\documentclass[preprint]{aastex}

\shortauthors{JONES, ET AL.}
\shorttitle{HIGHER RESOLUTION IMAGING OF NGC 4261} 

\begin{document}

\title{In the Shadow of the Accretion Disk: Higher\\ Resolution Imaging 
of the Central Parsec in NGC 4261} 

\author{Dayton L.~Jones} 
\affil{Jet Propulsion Laboratory, California Institute of Technology,
Mail Code 238-332,\\ 4800 Oak Grove Drive, Pasadena, CA 91109}
\email{dj@sgra.jpl.nasa.gov} 

\author{Ann E.~Wehrle}
\affil{Jet Propulsion Laboratory, California Institute of Technology, 
Mail Code 301-486,\\ 4800 Oak Grove Drive, Pasadena, CA 91109}
\email{aew@huey.jpl.nasa.gov} 

\author{B.~Glenn Piner} 
\affil{Jet Propulsion Laboratory, California Institute of Technology,
Mail Code 238-332,\\ 4800 Oak Grove Drive, Pasadena, CA 91109\\
and\\
Dept.~of Physics \& Astronomy, Whittier College, 13406 E.~Philadelphia 
Street,\\ Whittier, CA 90608}
\email{glenn@sgra.jpl.nasa.gov} 

\and

\author{David L.~Meier} 
\affil{Jet Propulsion Laboratory, California Institute of Technology,
Mail Code 238-332,\\ 4800 Oak Grove Drive, Pasadena, CA 91109}
\email{dlm@sgra.jpl.nasa.gov} 

\begin{abstract}  

The physical conditions in the inner parsec of accretion disks 
believed to orbit the central black holes in active galactic 
nuclei can be probed by imaging the absorption (by ionized gas
in the disk) of background emission from a radio counterjet.  
We report high angular resolution VLBI observations of the nearby 
($\sim 40$ Mpc) radio galaxy NGC 4261 that confirm 
free-free absorption of radio emission from a counterjet by a 
geometrically thin, nearly
edge-on disk at 1.6, 4.8, and 8.4 GHz.  The angular width and
depth of the absorption appears to increase with
decreasing frequency, as expected.  
We derive an average electron density of $\sim 10^{4}$
cm$^{-3}$ at a disk radius of about 0.2 pc, assuming that the 
inner disk inclination and opening angles are the same as at larger
radii.  Pressure balance between the thermal gas and the 
magnetic field in the disk implies an average field strength 
of $10^{-4}$ gauss at a radius of 0.2 pc.  
These are the closest-in free-free absorption measurements 
to date of the conditions 
in an extragalactic accretion disk orbiting a black hole with
a well-determined mass.  If a standard advection-dominated 
accretion flow exists in the disk center, then the transition
between thin and thick disk regions must occur at a radius 
less than 0.2 pc (4000 Schwarzschild radii).  

\end{abstract}


\section{Introduction}

The nearby low-luminosity (FR I) radio galaxy NGC 4261 (3C270) is a good 
candidate for the detection of free-free absorption by ionized gas in an 
inner accretion disk.  The galaxy is known to contain a central black hole
with a mass of $5 \times 10^{8} \ {\rm M}_{\odot}$ \citep[]{ffj96}, a nearly 
edge-on nuclear disk of gas and dust with a diameter of $\sim 100$ pc, and
a large-scale symmetric radio structure which implies that the radio
axis is close to the plane of the sky.  \citet[]{nolt93} gives a distance
of $27\ {\rm h}^{-1}$ Mpc, or 40 Mpc for an assumed Hubble constant
of $67 \ {\rm km\ s}^{-1}\ {\rm Mpc}^{-1}$.  At a distance of  
40 Mpc, 1 milliarcsecond (mas) corresponds to 0.2 pc.  

\begin{center}
\begin{figure}[h]
\vspace{65mm}
\includegraphics{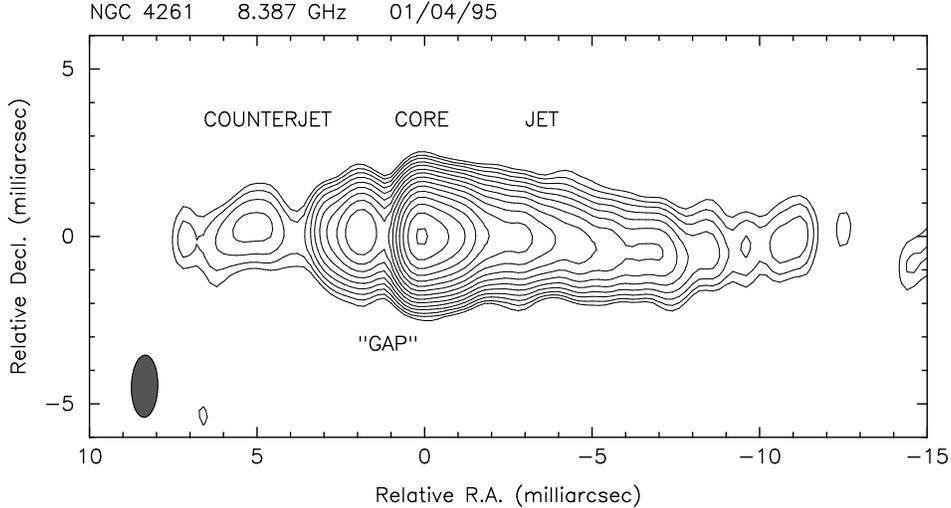}
\caption{VLBA image of NGC~4261 at 8.4 GHz.  The contours increase in steps 
of $\sqrt{2}$ starting at $\pm 0.75$\% of the peak, which is 99 mJy/beam. 
The restoring beam is $1.86 \times 0.79$ mas with major axis 
along position angle $-1.3^{\circ}$.
} \label{fig1}
\end{figure}
\end{center}

Previous VLBA 
observations of this galaxy revealed a parsec-scale radio jet and 
counterjet aligned with the kpc-scale jet (see Figure~\ref{fig1}; 
\citet[]{jw97}).
We believe the west-pointing jet is oriented slightly towards us 
(so the east-pointing jet is labeled the counterjet) based on the
orientation of the dust disk imaged by HST and the fact that the
west-pointing kpc-scale jet is slightly brighter over most of its
length in VLA images (\citet{bd85}; \citet{koff00}).  
The opening angle of the jets is less than $20^{\circ}$ during the
first 0.2 pc and less than $5^{\circ}$ during the first 0.8 pc.  At 8.4 GHz  
we found evidence for a narrow gap in radio brightness at the base of
the parsec-scale counterjet, just east (left) of the brightest peak 
which we identified as the core based on its inverted spectrum between
1.6 and 8.4 GHz \citep{jw97}.  We 
tentatively identified this gap as the signature of free-free absorption
by a nearly-edge on inner disk with a width much less than $0.1$ pc and an 
average electron density of $10^{3}-10^{8}\ {\rm cm}^{-3}$ over the 
inner 0.1 pc. 

The region labeled ``gap" in Figure~\ref{fig1} differs from other 
regions of reduced emission along the counterjet in two ways.  
First, it is uniquely deep and narrow.  This is difficult to see in
the contour maps (Figures~\ref{fig1} and \ref{fig7}), but is clear 
in the brightness profiles (Figures~\ref{fig11} and \ref{fig12}). 
Second, the ``gap" region has a strongly inverted radio  
spectrum, as shown in Figure~\ref{fig9}.  This is completely different 
from the spectrum anywhere else along the jet or counterjet.  
For these reasons we believe that the gap labeled in Figure~\ref{fig1} 
is not an intrinsically faint part of the counterjet but instead
is faint because of intervening absorption.   

\section{Observations}

We observed NGC 4261 at 1.6 and 4.9 GHz with the HALCA 
satellite and a ground array 
composed of 7 VLBA\footnote{ 
The Very Long Baseline Array (VLBA) and the Very Large Array (VLA) are
facilities of the National Radio Astronomy Observatory, which is operated
by Associated Universities, Inc., under a cooperative agreement with the 
National Science Foundation.}
antennas plus the 25-m antenna at Shanghai, China, the 34-m antenna
at Kashima, Japan, and the NASA Deep Space Network 70-m antenna at  
Tidbinbilla, Australia, at 1.6 GHz (22 June 1999) and 8 VLBA 
antennas plus the phased 27 antennas of the VLA at 4.9 GHz (27 June 1999). 
HALCA is an 8-m diameter antenna in Earth orbit, which is used in
combination with ground antennas by the VSOP project \citep[]{Hirax98}.
During both epochs the VLBA antennas at St.~Croix
and Hancock were unavailable, as was the North Liberty antenna 
during the 1.6 GHz session. 
Table~\ref{tab1} lists the antennas involved at each frequency. 
Data were recorded as two 16-MHz bandwidth channels with 
2-bit sampling by the Mark-III/VLBA systems and correlated at the VLBA
processor in Socorro.  At 1.6 GHz we recorded 1.634-1.666 GHz, and 
at 4.9 GHz we recorded 4.850-4.882 GHz. 
Both channels were sensitive to left circular
polarization.
The correlator produced a total of 32 spectral channels with an 
averaging time of 1 second to minimize time smearing of visibility
measurements on the ground-space baselines.

\begin{table}[ht]
\begin{center}
\caption{Antennas used at 1.6 and 4.9 GHz} \label{tab1}  
\vskip 12pt
\begin{tabular}{ccccc}
\hline \hline 
Antenna & Diameter & Typical~T$_{\rm SYS}$ & 1.6 GHz & 4.9 GHz \\
name & (m) & (K) & (Y/N) & (Y/N) \\
\hline 
VLBA-BR &  25 &  25-60 & Y & Y \\
VLBA-FD &  25 &  25-60 & Y & Y \\
VLBA-KP &  25 &  25-60 & Y & Y \\
VLBA-LA &  25 &  25-60 & Y & Y \\
VLBA-MK &  25 &  25-60 & Y & Y \\
VLBA-NL &  25 &  25-60 & N & Y \\ 
VLBA-OV &  25 &  25-60 & Y & Y \\
VLBA-PT &  25 &  25-60 & Y & Y \\
HALCA (VSOP)&  ~8 &  75-95 & N & Y \\
Kashima, Japan&  34 & \,$\approx 170$ & Y & N \\
Shanghai, China&  25 & \,$\approx 110$ & Y & N \\
Tidbinbilla, Aust.&70& $\sim 45$  & Y & N \\
\hline \hline
\end{tabular}
\end{center}

\end{table}

Fringe-fitting was carried out in AIPS after applying {\it a priori}
amplitude calibration.  For VLBA antennas we used continuously measured  
system temperatures, while for the phased VLA we used 
measured ${\rm T}_{\rm A} /
{\rm T}_{\rm SYS}$ values with an assumed source flux density of 5 Jy. 
The remaining antennas did not provide real-time calibration data, so
we used typical gain and system temperature
values obtained from the VSOP web site.  Fringes were found to all
antennas at 1.6 GHz except HALCA.  The 
{\it a priori} amplitude calibration for Tidbinbilla was dramatically
incorrect (by factors of 13.3 and 13.7 in the two channels) 
for unknown reasons.  We calibrated Tidbinbilla by imaging
the compact structure of the source using VLBA data, then holding the
VLBA antenna gains fixed and allowing the Tidbinbilla gain to vary. 
A single time-independent gain correction was determined in this way.
This produced a good match in correlated flux density where the
projected VLBA and Tidbinbilla baselines overlap.  At 4.9 GHz fringes
were found to all antennas, including HALCA.  Similar corrections 
to the {\it a priori} amplitude calibration for HALCA and the phased 
VLA were applied.  For HALCA these corrections were very large 
($\sim 100$) 
while for the phased VLA they were 0.24 and 0.25 in the two channels. 
The reason for the large HALCA amplitude correction is not known, but 
the VLA corrections may have resulted from our assumed total flux
density for NGC 4261.  
After correction of the HALCA fringe amplitudes, the HALCA-ground baselines
agreed well with ground-ground baselines and the observed SNR
was consistent with the expected noise levels.  This was true for 
the baseline phases as well, which were unaffected by the amplitude 
corrections.  

In both observations we found that averaging in frequency over both 
16-MHz channels in AIPS (using task AVSPC)  
produced large, baseline-dependent amplitude
reductions even though the post-fringe-fit visibility phases were 
flat and continuous between channels.  Averaging over frequency within
each 16-MHz band separately fixed this problem.  The Caltech program
Difmap \citep[]{spt94} was 
used for detailed data editing, self-calibration, and image
deconvolution.  Both 16-MHz bands were combined by Difmap during imaging.

Imaging within Difmap used uniform weighting with the weight of HALCA
data increased by a factor of 500.  This prevented the intrinsically
more sensitive ground baselines from completely determining the
image resolution.  
Several iterations of phase-only
self calibration, followed by amplitude self calibration iterations 
with decreasing time scales, resulted in good fits ($\chi^{2} \approx
1$) between the source model and the data. 
The restoring beam at 1.6 GHz was $10.3 \times 2.2$ mas, in position
angle -45$^{\circ}$.  The restoring beam at 4.9 GHz was $2.8 \times
1.1$ mas, in position angle -5$^{\circ}$.  The off-source rms
noise levels in the images were 0.1 and 0.3 mJy/beam at 1.6 and 
4.9 GHz respectively.

\section{Results}

\subsection{1.6 GHz Image} 

\begin{center}
\begin{figure}[hb]
\vspace{65mm}
\includegraphics{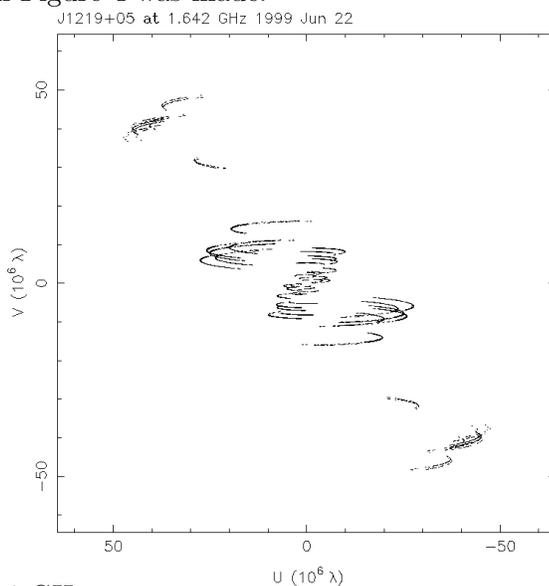}
\caption{(u,v) coverage at 1.6 GHz. 
} \label{fig2}
\end{figure}
\end{center}

\begin{center}
\begin{figure}
\vspace{70mm}
\includegraphics{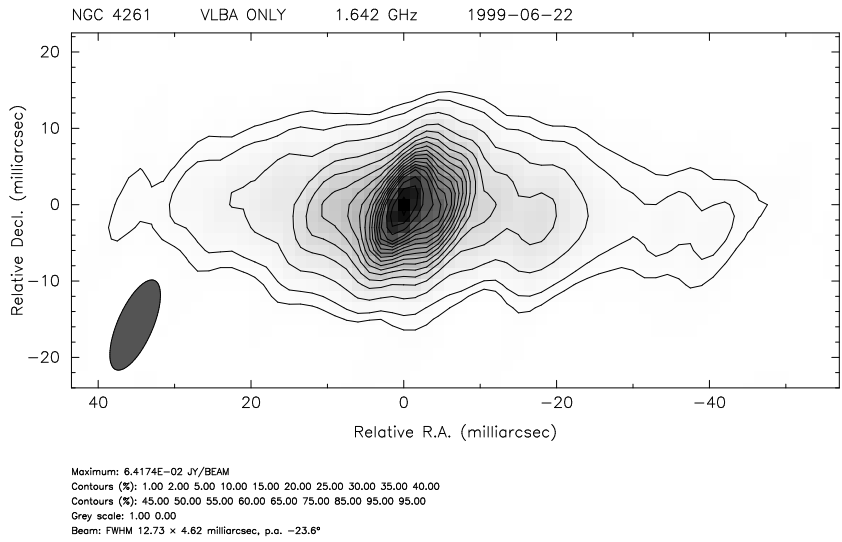}
\caption{VLBA image of NGC~4261 at 1.6 GHz.  The contours are 1, 2,
and 5\% followed by steps of 5\% of the peak (64 mJy/beam).  The 
restoring beam is $12.73 \times 4.62$ mas with major axis along
position angle $-23.6^{\circ}$. 
} \label{fig3}
\end{figure}
\end{center}

\begin{center}
\begin{figure}
\vspace{70mm}
\includegraphics{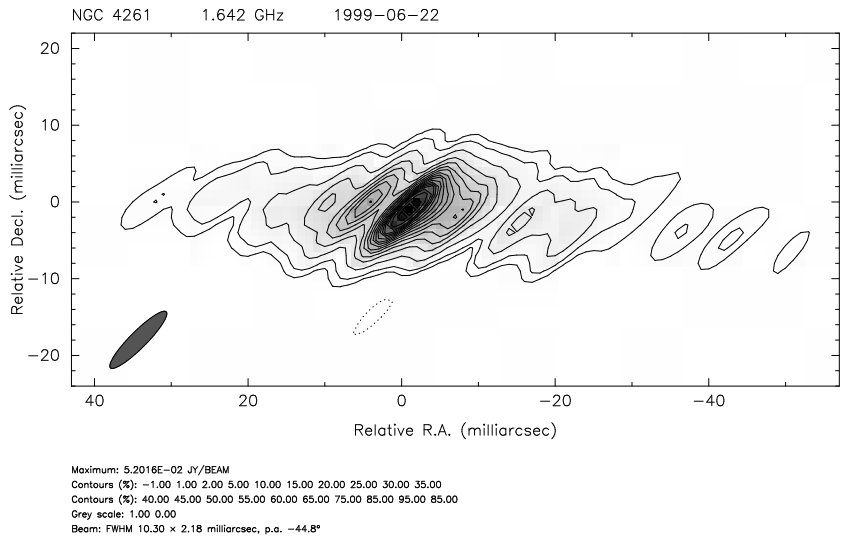}
\caption{VLBA+Tidbinbilla image of NGC~4261 at 1.6 GHz. 
Note the asymmetry near the core, which is not visible with 
the lower angular resolution in Figure~\ref{fig3}.  
The coutours are the same percentages of the peak (52 mJy/beam)  
as in Figure~\ref{fig3}, and the restoring beam is $10.30 
\times 2.18$ mas with major axis along position angle $-44.8^{\circ}$. 
} \label{fig4}
\end{figure}
\end{center}

\begin{center}
\begin{figure}[ht]
\vspace{95mm}
\includegraphics{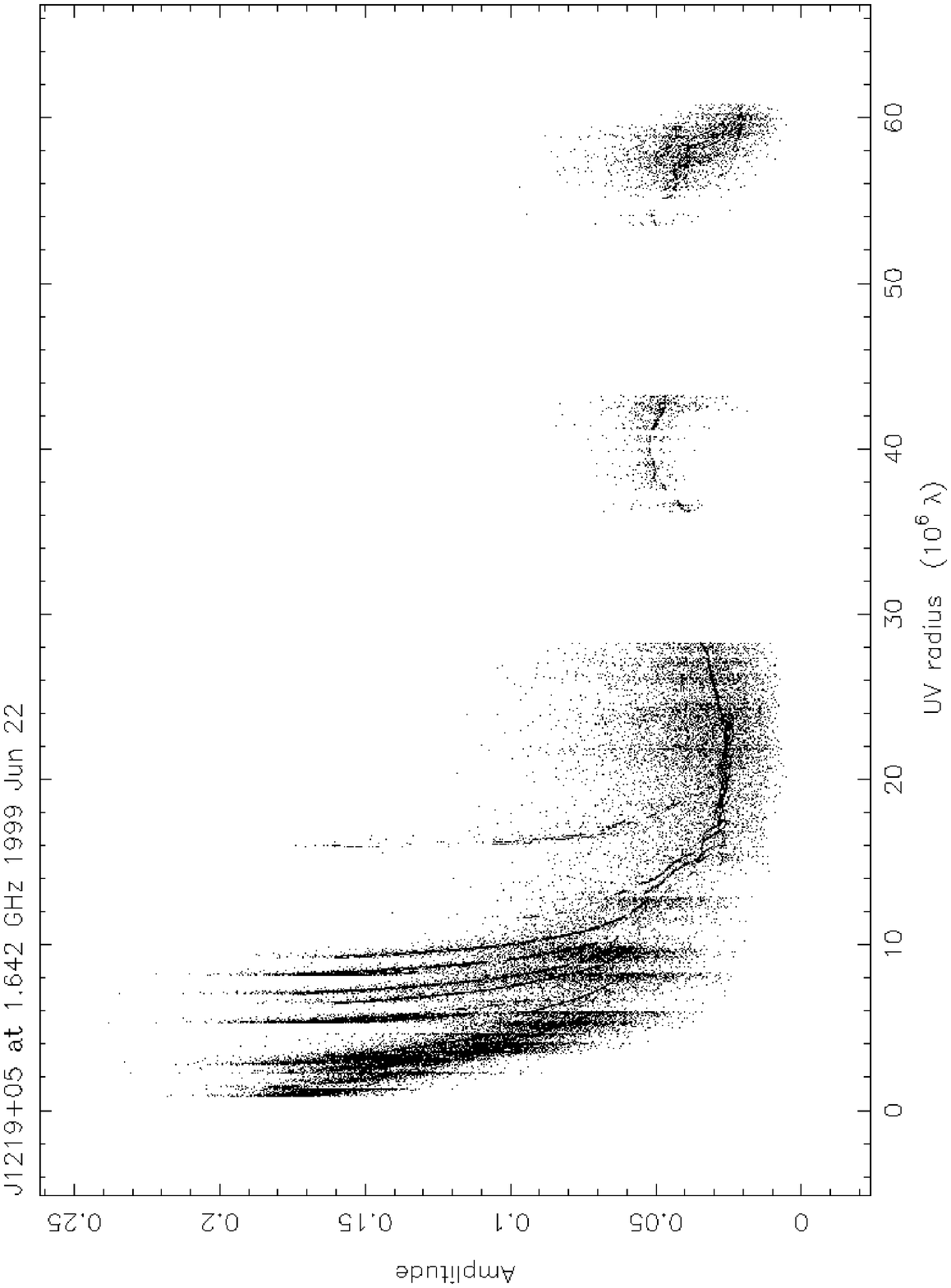}
\caption{Visibility amplitude (in Jy) as a function of projected
baseline length at 1.6 GHz. 
} \label{fig5}
\end{figure}
\end{center}

Our image at 1.6 GHz has 
more than twice the angular resolution of our previous 
1.6 GHz image \citep[]{jw97} due to the addition of Tidbinbilla,
even though no HALCA data is included 
(see Figure~\ref{fig2}).
The previous image showed a symmetric structure, with the jet and
counterjet extending west and east from the core.  An image made
from our new 1.6 GHz data, but using only VLBA antennas, also shows
a largely symmetric structure (Figure~\ref{fig3}).  No evidence
for absorption is seen in this image.  However, with the higher 
resolution provided by adding Tidbinbilla to the VLBA  
we do detect a narrow gap in emission just east of the core, at the
base of the counterjet (Figure~\ref{fig4}).  
The width of the gap is less than 2 mas.
Figure~\ref{fig5} shows the visibility data from which the image 
in Figure~\ref{fig4} was made.

\subsection{4.9 GHz Image}

\begin{center}
\begin{figure}
\vspace{70mm}
\includegraphics{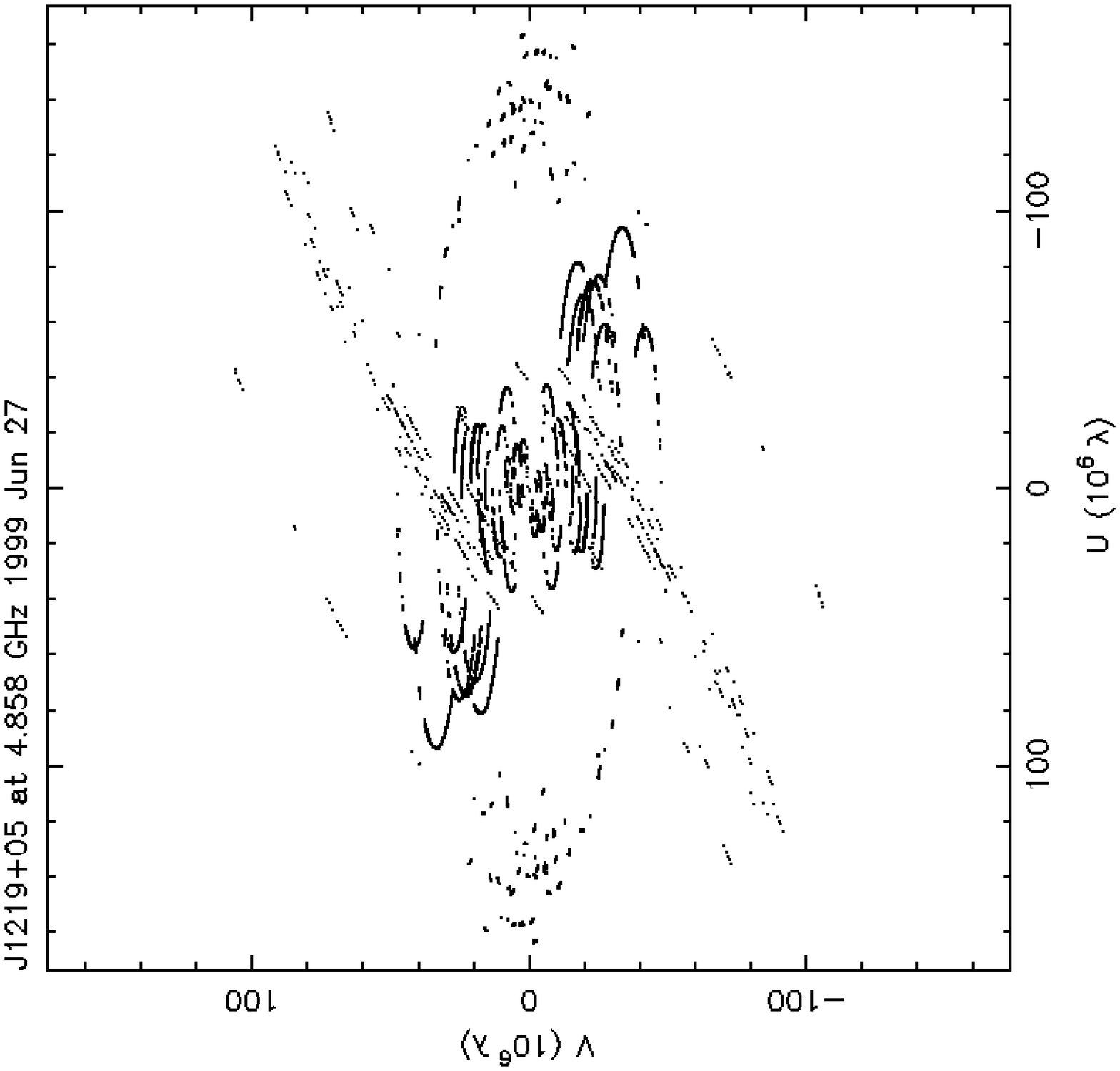}
\caption{(u,v) coverage at 4.9 GHz. 
} \label{fig6}
\end{figure}
\end{center}

\begin{center}
\begin{figure}
\vspace{70mm}
\includegraphics{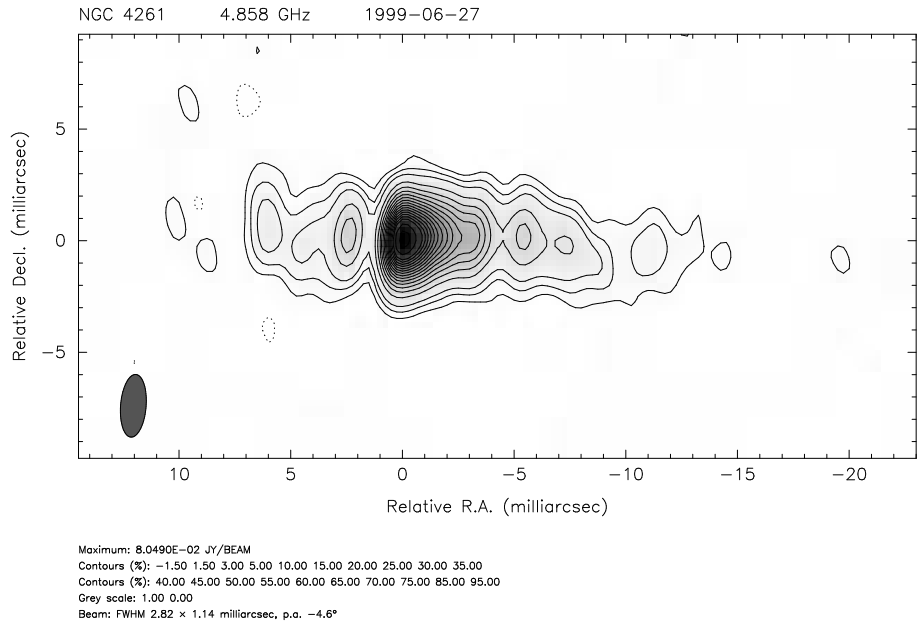}
\caption{Full resolution (VLBA+HALCA) image of NGC~4261 at
4.9 GHz.  The contours are -1, 1, 2, 4, 8, 16, 32, and 64\% of
the peak (165 mJy/beam), and the restoring beam is $1.06 \times 0.29$
mas with the major axis along position angle -18.3$^{\circ}$.
The total flux density in the image is 388 mJy. 
} \label{fig7}
\end{figure}
\end{center}

We detected fringes to HALCA at 4.9 GHz only when the projected Earth-space
baselines were less than one Earth diameter.  However, the HALCA data fills 
in the (u,v) coverage hole between continental VLBA baselines and
those to Mauna Kea, and also increases the north-south resolution
by a factor of two (see Figure~\ref{fig6}).  
Our 4.9 GHz image is shown in Figure~\ref{fig7}.
Note that the gap in emission is again seen just east of the
peak.  A careful comparison of brightness along the radio axis
at 4.9 and 8.4 GHz shows that the gap is both deeper and wider
at 4.9 GHz, as expected from free-free absorption.  The region of
the gap contains far less flux at 4.9 GHz than at 8.4 GHz, and
thus has a very inverted spectrum.  The brightest peak (core)
has a slightly less inverted spectrum, and the distant parts of
both the jet and counterjet have generally steep (normal) spectra.
We define the spectral index $\alpha$ by 
${{\rm S}_{\nu}} \propto {{\nu}^{\alpha}}$.  Thus, an ``inverted"
spectrum means $\alpha > 0$, characteristic of either synchrotron
self-absorption or free-free absorption, and a ``steep" spectrum means 
$\alpha < 0$, characteristic of optically thin synchrotron emission.  

\begin{center}
\begin{figure}
\vspace{70mm}
\includegraphics{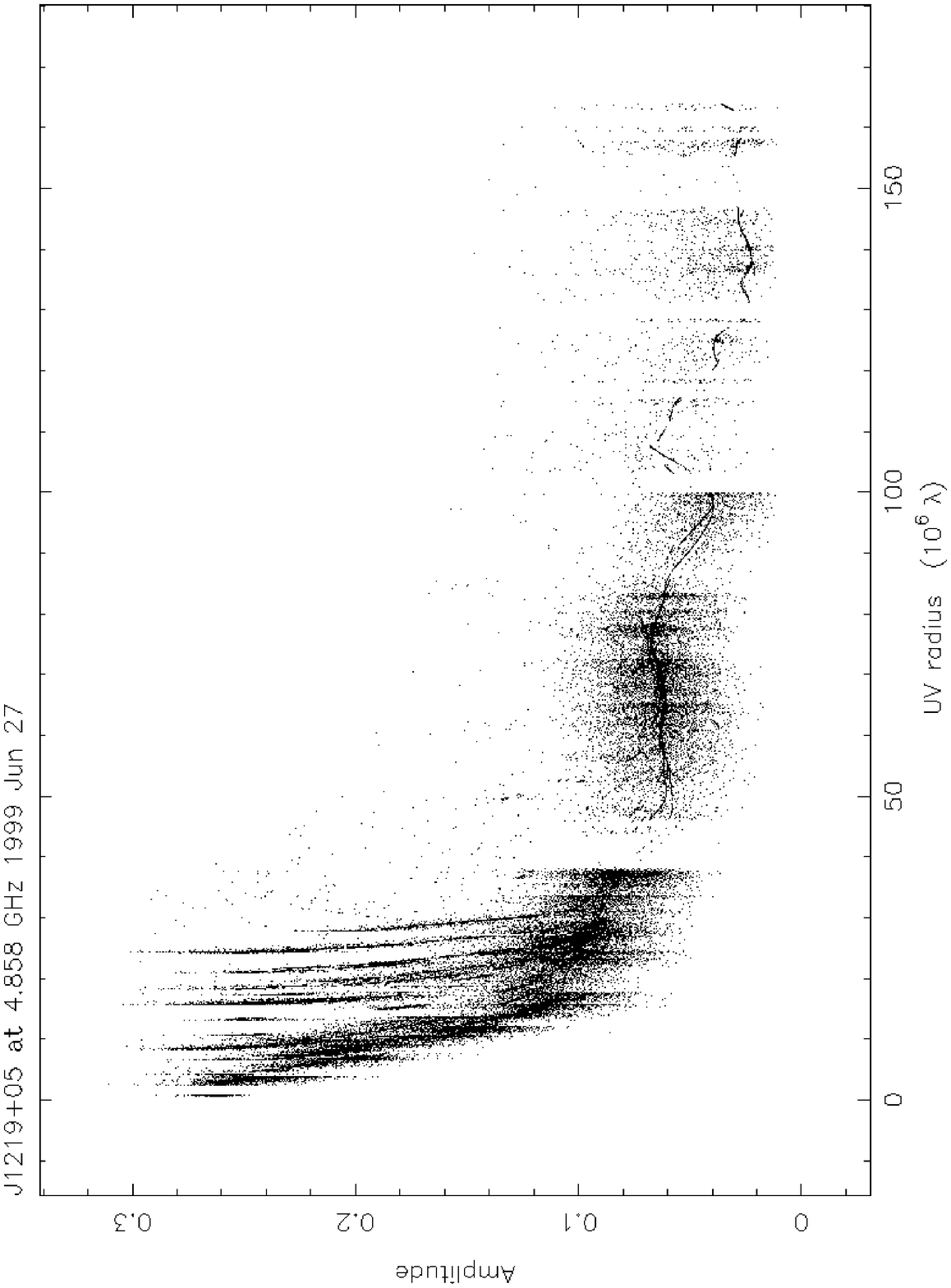} 
\caption{Visibility amplitude (in Jy) as a function of projected
baseline length at 4.9 GHz.  
} \label{fig8}
\end{figure}
\end{center}

Figure~\ref{fig8} shows the visibility data from which the image
in Figure~\ref{fig7} was made, showing significant structure 
on the longest baselines.

\newpage

\subsection{Spectral Index Distribution}

Accurate maps of spectral index require images with matched 
angular resolutions at two or more frequencies and accurate
registration of the images.  The use of HALCA at 4.9 GHz provided
east-west resolution similar to that obtained with the VLBA alone
at 8.4 GHz.  For registration we used the fact that the core-gap
angular separation is nearly identical at both frequencies (1.2
and 1.1 mas at 4.9 and 8.4 GHz) and that this separation has
remained unchanged during three epochs spanning 4 years at 
8.4 GHz \citep[]{pjw00}.  This indicates that the core (base of the 
jet) and gap (absorption by disk) are features which remain fixed in
position over 4 years, and, consequently, they can be used to align
VLBI images with similar resolutions.  Our three epochs of data
at 8.4 GHz show no evidence for large changes in flux density  
over 4 years.  Consequently, we believe that 
the 4 month difference in epoch between our 4.9 and 8.4 GHz
images will not lead to significant errors in spectral index
determinations.  

\begin{center}
\begin{figure}
\vspace{68mm}
\includegraphics{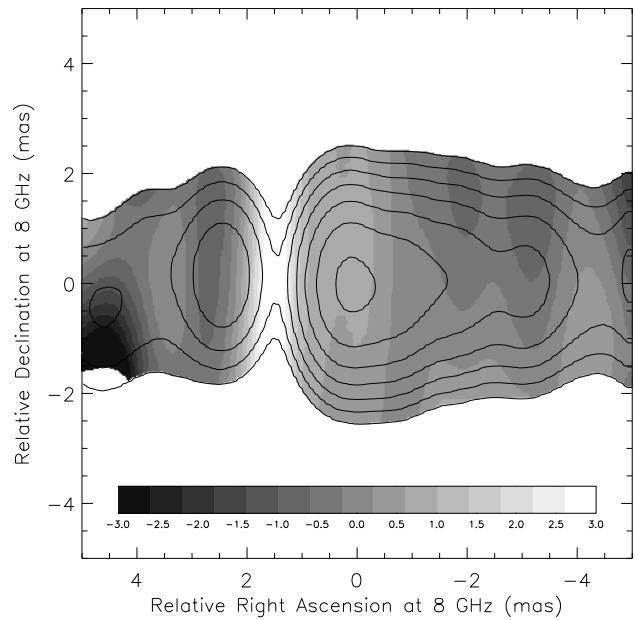}
\caption{Spectral index map of NGC~4261 made by combining 
images (without residuals) at 4.9 GHz and 8.4 GHz. 
} \label{fig9}
\end{figure}
\end{center}

Figure~\ref{fig9} shows the spectral index map produced from
our matched-resolution images at 4.8 and 8.4 GHz, without residuals.
The registration of non-phase-referenced VLBI images is arbitrary,
but we can use a combination of four features -- the peak, gap,
counterjet peak just east of the gap, and the jet feature 11 mas
west of the peak.  A single offset allowed all four of these features
to be matched in position between the images.  

\section{Discussion}

\subsection{Jet Opening Angle}

We see no evidence for an initial large jet opening angle similar
to that described by \citet{jbl99} for M87.  However, we are not
probing as close to the base of the jets in terms of gravitational
radii because M87 is closer than NGC 4261 and the central black
hole in M87 is more massive. 
The parsec-scale jet  
and counterjet in NGC 4261 are unresolved in the transverse-to-the-jet 
direction at all of our observing frequencies.  
The core component
in Figure~\ref{fig1} must be larger than $\sim 0.04$ mas 
($\sim 9$ light days) in diameter
to avoid synchrotron self-absorption, but the inverted spectrum
of the core in Figure~\ref{fig9} suggests that the core may
be partially self-absorbed.  The largest opening angle comes
from assuming that the base of the jet (core) has zero angular
size.  Combined with our upper
limit of $\sim 1$ mas for the jet transverse width at 10 mas from 
the core (as far out as we can detect the jet with high SNR at
4.9 or 8.4 GHz)   
this gives an upper limit of $\sim 6^{\circ}$ for the full jet
opening angle over the first 2 pc.  This assumes the jet is conical
and in the
plane of the sky; a more end-on view would reduce our upper limit
for the jet opening angle.

A lower limit for the brightness temperature of the core 
is $\approx 10^{10}$ K.  Although we can not rule out a Doppler
factor significantly greater than unity, there is no strong
evidence (rapid variability, large jet/counterjet brightness ratio,
or large proper motions) 
that would suggest a large Doppler factor in this source.

\subsection{Disk Structure}

The large-scale HST disk, if intrinsically circular, is oriented
$64^{\circ}$ from face-on and is $14^{\circ}$ from being perpendicular
to the radio axis in the plane of the sky (\citet{ffj96}; \citet{jaf96}).
If we assume that $14^{\circ} \ \times \ {\rm sin}(64^{\circ}) = 
12.6^{\circ}$ is 
typical of the offset between 
the plane of the disk and the radio axis, the radio axis is likely
oriented between $51^{\circ}$ and $77^{\circ}$ away from our line of sight.

VLBI observations by \citet{vanL00} have detected absorption by 
neutral hydrogen (HI) as  
close as $\sim 3$ pc from the center of NGC 4261.  HI absorption has also
been seen close to the center of NGC 1052 by \citet{k99}, indicating
that the situation in NGC 4261 is not unique.
\citet{vanL00} find an HI column density of $2.5 \times 10^{19} \  
{\rm T}_{\rm SPIN} \ {\rm cm}^{-2}$ in front of the counterjet, 
where T$_{\rm SPIN}$ is the spin temperature.  These authors deduce
that the HI is in a thin disk with a full opening angle of $13^{\circ}$. 
This implies an average HI density of $6 \times 10^{2} \ {\rm cm}^{-3}$
for T$_{\rm SPIN} = 100$ K.  This is a lower limit because T$_{\rm SPIN}$
may be higher in this region and the filling factor may be smaller than
unity.  The disk opening angle found by 
\citet{vanL00} is comparable to the upper limit of \citet{ccc99}, who  
deduce a typical disk thickness/size ($h/r$) ratio of $\le 0.15$ for their 
sample of low-luminosity radio galaxies.  This sample includes NGC 4261. 
The innermost part of the disk could be extremely thin.  For the
well-studied disk in the peculiar Seyfert galaxy NGC 4258, where 
water masers provide powerful  
constraints on the disk geometry, the inner 0.25 pc of the maser
disk has a height/radius ratio $h/r \approx 0.01$ \citep{moran00}. 

\begin{center}
\begin{figure}
\vspace{135mm}
\includegraphics{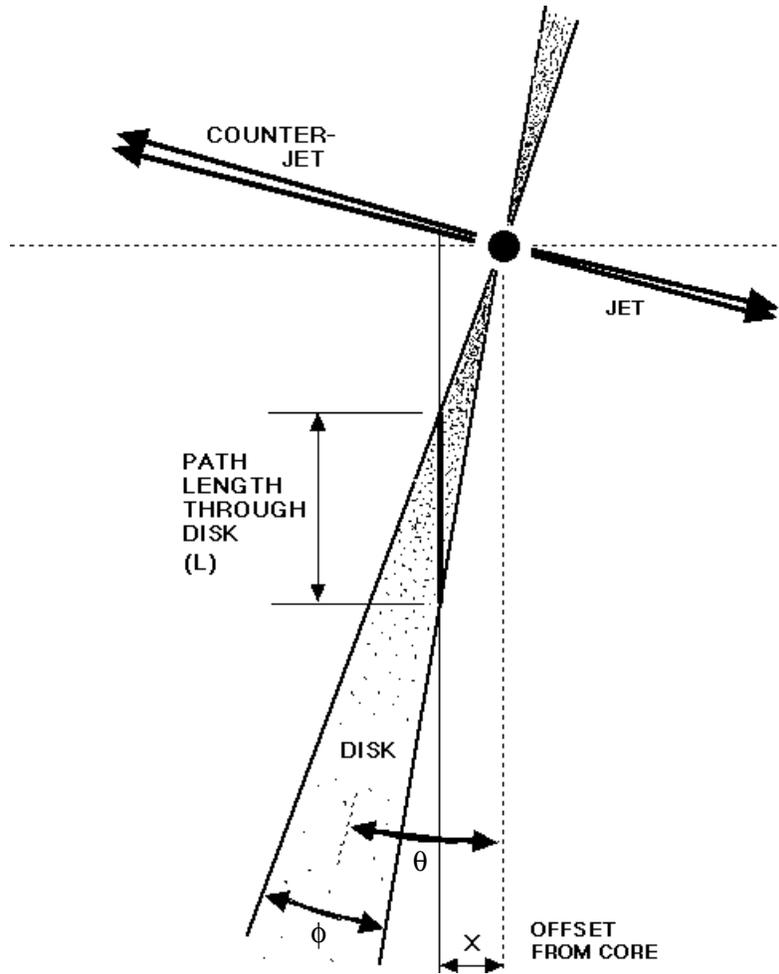}
\caption{Possible geometry of inner few parsecs in NGC 4261,
showing the radio jet and counterjet, location of the central
black hole, and the inner part of a geometrically thin
accretion disk.  The angle $\phi$ is the full opening
angle of the disk, and $\theta$ is the angle between the
plane of the disk and our line of sight ($\theta \ = \ 
90^{\circ}-{\rm inclination}$ angle of disk $\approx$  
angle between radio axis and the plane of the sky).
} \label{fig10}
\end{figure}
\end{center}

For a $26^{\circ}$ angle between the plane of the disk and our
line of sight, the angular width of absorption in Figures~\ref{fig4}, 
\ref{fig7}, and \ref{fig1} implies that the optical depth becomes $<< 1$ 
at a disk radius $r > 1.0$ pc at 1.6 GHz, $r > 0.6$ pc at 4.9 GHz, and 
and $r > 0.5$ pc at 8.4 GHz. 
Using the \citet{vanL00} disk opening angle of $13^{\circ}$ ($h/r =
0.22$) we get disk thicknesses of 0.22 pc at a radius of 1.0 pc, 
0.13 pc at a radius of 0.6 pc, and 0.11 pc at a radius of 0.5 pc.
The line-of-sight path lengths 
through the disk are $L = 0.52$ pc, $L = 0.31$ pc, and $L = 0.29$ pc 
respectively (see Figure~\ref{fig10}, with $\phi = 13^{\circ}$ 
and $\theta = 26^{\circ}$).
Alternatively, using the Chiaberg, et al., value of $h/r \le 0.15$ 
(for which $\phi < 9^{\circ}$ in Figure~\ref{fig10}) we get upper
limits for the disk thickness of 0.15 pc at a radius of 1.0 pc, 
0.09 pc at a radius of 0.6 pc, and 0.08 pc at a radius of 0.5 pc.  
The resulting path lengths through the disk are $L < 0.36$ pc, $L < 
0.22$ pc, and $L < 0.20$ pc at radii of 1.0, 0.6, and 0.5 pc.
Finally, using the disk model in the appendix of 
\citet{jones00} we get a disk thickness of 0.012 pc, 
0.006 pc, and 0.004 pc at radii of 1.0 pc, 0.6 pc, and 0.5 pc,
respectively.  The corresponding values of $\phi$ in 
Figure~\ref{fig10} are $0.7^{\circ}$, $0.6^{\circ}$, 
and $0.5^{\circ}$.  
These values give path lengths through the disk of 0.03 pc, 
0.014 pc, and 0.011 pc at 1.6, 4.9, and 8.4 GHz. 
The path lengths through the disk for each of the three disk
models just discussed are summarized in Table~\ref{tab2}. 

\begin{table}[htbp]
\begin{center}
\vspace{45mm}
\caption{Line-of-Sight Path Lengths through Disk} \label{tab2}
\vskip 12pt
\begin{tabular}{lcccc}
\hline \hline 
Disk Model & Angles (see Figure~\ref{fig10}) & 1.6 GHz & 4.9 GHz & 8.4 GHz \\
\hline 
van Langevelde, et al. & $\theta = 26^{\circ}, \ \phi = 13^{\circ}$ & 0.52 pc & 0.31 pc & 0.29 pc \\
Chiaberg, et al. & $\theta = 26^{\circ}, \ \phi < 9^{\circ}$ & $<$ 0.36 pc & $<$ 0.22 pc & $<$ 0.20 pc \\
Jones, et al. & $\theta = 26^{\circ}, \ \phi = 0.5-0.7^{\circ}$ & 0.03 pc & 0.014 pc & 0.011 pc \\
\hline \hline
\end{tabular}
\end{center}

\end{table}

Therefore our choice of disk opening angle can change the
line-of-sight path lengths through the disk by 
factors of 17, 22, and 26 at 1.6, 4.9, and 8.4 GHz.  However,
for a given observed optical depth the average electron density
depends on $\sqrt{L}$, so our derived values of $n_{e}$ will
vary only by factors of 4 or 5 if we 
assume different disk opening angles.  

It is quite possible that the central region of this 
low-luminosity radio galaxy contains an  
advection-dominated accretion flow (ADAF).  However, because we
observe a thin disk down to a radius of 0.2 pc ($\sim 4000$
Schwarzschild radii), we conclude that the transition radius 
between the outer thin disk and an inner ADAF must be smaller
than this value.  This is a useful constraint (see, {\it e.g.},
\citet{nmq98}) and is consistent with the upper limit for ADAF
size of $\sim 100$ Schwarzschild radii in NGC 4258 \citep{hgm98}. 
Of course, the relatively cool thin disk could contain a
non-standard, hot, advection-dominated corona; our observations
would not have been sensitive to it.

\subsection{Optical Depth and Electron Density}

\begin{center}
\begin{figure} 
\vspace{75mm}
\includegraphics{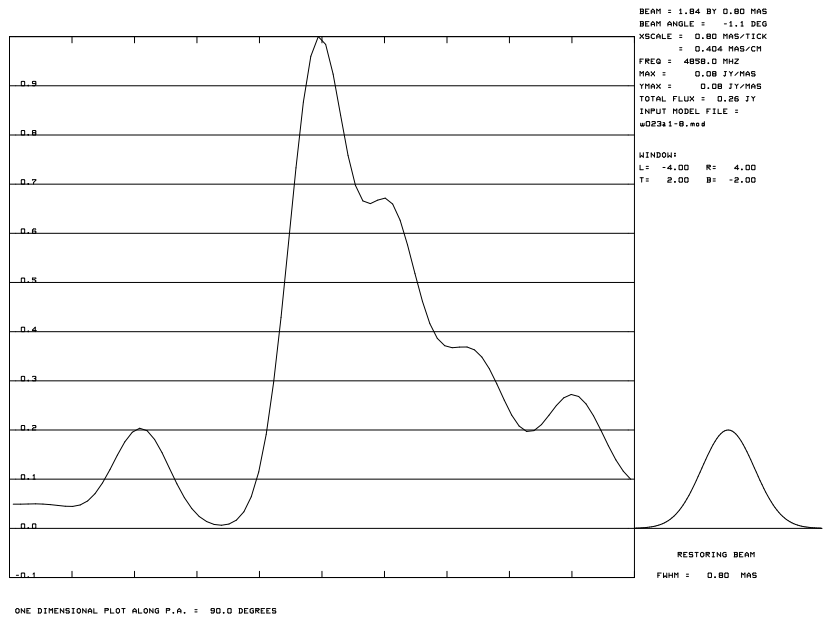}
\caption{Brightness profile along the radio axis of  
NGC~4261 at 4.9 GHz.   
} \label{fig11}
\end{figure}
\end{center}

\begin{center}
\begin{figure}
\vspace{68mm}
\includegraphics{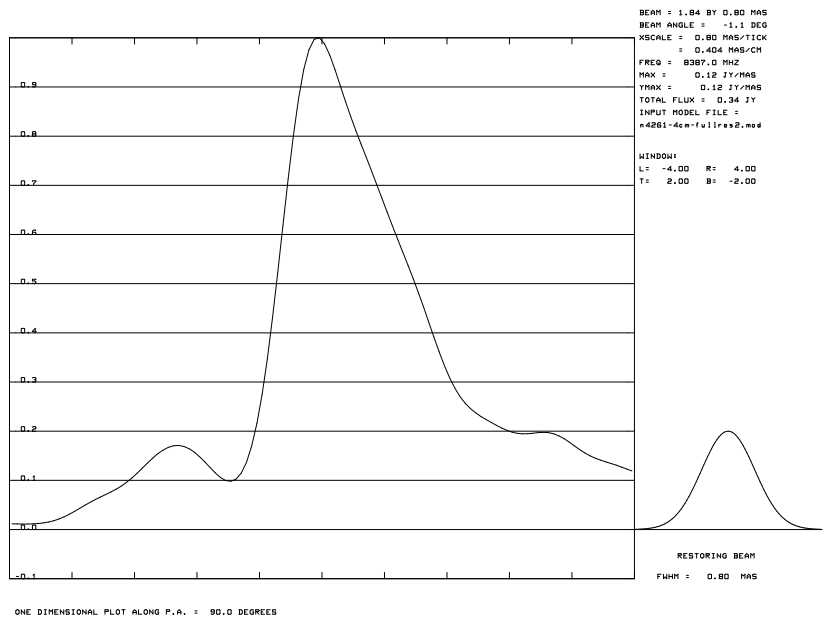}
\caption{Brightness profile along the radio axis of
NGC~4261 at 8.4 GHz, convolved with the same beam as in
Figure~\ref{fig11}. 
} \label{fig12}
\end{figure}
\end{center}

If we combine the disk opening angle of $\phi = 13^{\circ}$ from 
\citet{vanL00} with the disk orientation angle  
of $\theta = 26^{\circ}$ from \citet{ffj96} and \citet{jaf96}   
we find that the line of sight through the disk 
is ${x\,[{\rm cot}(19.5^{\circ}) - {\rm cot}(32.5^{\circ})]} 
\approx 1.3\,x$, where $x$ is the projected distance on the sky 
between the center of the disk (core) and the line of sight
(see Figure~\ref{fig10}).  
From figures~\ref{fig11} and \ref{fig12} we see that the location
of the absorption (gap) is about 1.2 mas from the peak at 4.9 GHz
and 1.1 mas from the peak at 8.4 GHz.   
An angular distance of 1.2 mas (0.24 pc) between the core and 
the center of the absorption gives a
total path length through the disk of 
0.31 pc.  For $x = 1.1$ mas (0.22 pc), the line-of-sight 
path length $L = 0.29$ pc.
These are sufficiently close that we will assume equal path
lengths at both frequencies. 

We would like to determine the radial distribution of electron
density in the accretion disk, but this requires measurement 
of the optical depth along multiple paths that pass through the
disk at different radii.  The absorption we see at 4.9 and 8.4
GHz occurs in nearly the same narrow region.  At 1.6 GHz the 
absorption is presumably larger, but the absorbed region is 
only partially resolved with our angular resolution so we have
only a lower limit for the optical depth.  A more sensitive 1.6 
GHz experiment (that might allow fringes to HALCA to be detected)
or a global VLBI experiment at 2.3 GHz could help.  However, the
nearly edge-on orientation of the inner disk makes any 
determination of spatial variations in optical depth difficult. 

Assuming an intrinsic spectral index of $\alpha$ and a covering
factor of unity, the observed brightness ratio at two frequencies
$\nu_1$ and $\nu_2$ is given by $({{\nu_1}/{\nu_2}})^{\alpha}\  
e^{-({\tau_1}-{\tau_2})}$.  
The apparent reduction in brightness
is about a factor of 30 at 4.9 GHz and a factor of 4 at 8.4 GHz. 
This implies that $\tau_{4.9} \approx 3.4$ and $\tau_{8.4} 
\approx 1.4$, so ${{\tau_{4.9}}-{\tau_{8.4}}} \approx 2.0$.  
As a check on the plausibility of these values, 
we note that a brightness ratio of 7.5 between 8.4 and 4.9 GHz
implies that ${{\tau_{4.9}}-{\tau_{8.4}}} \ = \ 
{\rm ln}[7.5\ ({4.9/8.4})^{-0.3}] \approx 2.2$.  
The spectral index $\alpha = -0.3$ comes from \citet{pjw00}.
The prediction of this simple formula for ${\tau_1}-{\tau_2}$ 
(= 2.2) is in  
reasonable agreement with the measured value of 2.0, considering
that we have not taken  
into account the (small) difference in path length through
the disk at the two frequencies.  

The free-free optical depth is given by
\citet{walker00} as 

$$\tau = {9.8 \times 10^{-3}}\ {L}\ {{n_e}^2}\ {T^{-1.5}}\ 
{\nu^{-2}}\ [17.7 + {\rm ln}({T^{1.5}}\ \nu^{-1})]$$

where the path length $L$ is in cm (not pc as in \citet{walker00};
see \citet{pach70}), $n_e$ 
is in electrons cm$^{-3}$, $T$ is in K, and $\nu$ is
in Hz.  

Astrophysical plasmas undergoing sufficient heating to stay
ionized generally occupy one of two phases: a very hot phase
that fills the local space at roughly the virial temperature 
or a warm phase at about $10^{4}$K.  (Under ambient conditions,
the latter phase maintains pressure balance with the former
phase, but in a black hole accretion disk situation the 
cooler gas will collect in a Keplerian disk whose pressure
and density are determined by the black hole's tidal forces.) 
The virial temperature at $\sim 0.3$ pc from a $5 \times 10^{8} \, 
M_{\odot}$ black hole is nearly $10^{9}$ K, producing a very
tenuous, spherical plasma with little free-free absorption, 
rather than a thin disk with substantial absorption.  A 
spherical, low-absorption plasma is inconsistent with our
detection of narrow, deep absorption features.
We therefore choose the warm, efficiently-radiating $10^{4}$ K 
accretion phase, for which we have developed a simple model in 
\citet{jones00}. 
In this case the Gaunt factor
(in square brackets above) for gas within the disk  
is approximately 9 at both 4.9  
and 8.4 GHz.  We now have estimates for $\tau,\ T_{e},\  
{\rm and}\ L$, and can calculate the average $n_e$ along 
the line of sight for each value of $\nu$:  

$${n_e} = 3.4\ {\tau^{1/2}}\ L^{-1/2}\ \nu\ {T^{3/4}}.$$ 

At $\nu = 4.9 \times 10^{9}$ Hz, we have $\tau \approx 3.4$
and $L \approx 0.31\ {\rm pc} \approx 9 \times 10^{17}$ cm.  
Adding the assumption of ${T_e} \approx
10^{4}$ gives us ${n_e} \sim 3 \times 10^{4}$ cm$^{-3}$.  
At 8.4 GHz we get ${n_e} \sim 2 \times 10^{4}$ cm$^{-3}$.  
Given the uncertainties in the optical depths and path 
lengths, these values for $n_e$ are not significantly different.
These values are consistent with the electron densities derived by
\citet{jones00} based on lower resolution images.  Note that
if the thinner model disk from \citet{jones00} is used the
average electron densities would be only 4-5 times higher. 
In all cases a shorter path length or higher gas   
temperature would increase the average $n_e$.

The electron densities derived here are low enough that Thompson
scattering should not be a significant source of additional 
absorption.  The Thompson scattering coefficient per free 
electron is $6.7 \times 10^{-25}\ {\rm cm}^{-2}$, so   
a column density of $\sim 5 \times 10^{22}$ electrons cm$^{-2}$
gives us an absorption of only $\sim 10^{-3}$ due to Thompson
scattering. 
The plasma oscillation frequency is given by $9 \times 10^{3} \ 
{{n_e}^{1/2}}$ Hz, which is $< 2$ MHz for ${n_e} \approx
2 \times 10^{4}$.  This is a factor of $\sim 10^{3}$ below
our observing frequencies and therefore we expect refraction
or reflection of radiation within the disk plasma to 
be negligible. 

Our result can be compared with that obtained by \citet[]{walker00}
for 3C84.  These authors find ${n_e} \approx 2 \times 10^{4} \ 
{\rm cm}^{-3}$ for a 1 pc path length, assuming ${T_e} \approx 
10^{4}$ K.  We find very similar average electron densities 
for the inner pc of the disk in NGC 4261, whose low frequency
radio luminosity is about 7 times smaller than for 3C84 
\citep[]{ss80}, suggesting that the
density of plasma in accretion disks does not depend strongly
on the radio luminosity of the central engine.  
It should be noted that very small or very large optical 
depths are difficult to measure accurately, and if typical 
path lengths through the ionized absorbing medium are 
similar in different galaxies this observational selection
effect will bias the distribution of derived electron densities
toward similar values.  
In any case, a larger number of sources covering a larger range 
of radio luminosities will need to be compared in this way 
before firm conclusions can be made.

\subsection{Disk Magnetic Field} 

If the accretion-generated luminosity is less than $10^{-3}$ 
times the Eddington luminosiity, there will be no 
radiation-pressure dominated region in the accretion disk (\citet[]{ss73}; 
\citet[]{m99}).  This may be the case in NGC 4261,
whose nuclear luminosity is quite small for an active galaxy. 
If this is the case, the 
average magnetic field $B$ in the disk can be estimated by 
assuming pressure balance between the magnetic field and the 
thermal disk gas: $B \approx \sqrt{8 \pi {\alpha_{\rm v}} 
{n_e} k {T_e}}$, where here $\alpha_{\rm v}$ is the \citet{ss73} 
viscosity parameter and $k$ is the Boltzmann constant.   
For $\alpha_{\rm v} = 0.01$, ${n_e} = 2 \times 10^{4} \ {\rm cm}^{-3}$,
and $T_{e} = 10^{4}\,{\rm K}$ 
we get $B \approx 10^{-4}$ gauss.  This
applies at disk radii near 0.2 pc. 
There is considerable uncertainty in these parameter values, 
particularly in the assumed value of $\alpha_{\rm v}$.  

We can calculate the expected rotation measure from $ 2.7
\times 10^{-13}\, {n_e}\, L\, B$, where $L$ is the path length
(still in cm) 
and $B$ is the parallel component of the magnetic field. 
Using the magnetic field and electron density derived above 
we get rotation measures of about $6 \times 10^{5}$ 
radians m$^{-2}$ at 4.8 GHz and $4 \times 10^{5}$ 
radians m$^{-2}$ at 8.4 GHz.  
These are actually upper limits because they assume that
the magnetic field is ordered and parallel to our line 
of sight, which is unlikely.  
In fact, shear created by differential rotation within 
the accretion disk may cause the magnetic field to be 
aligned along the direction of gas rotation.  This 
would mean that the magnetic field was mainly orthogonal  
to our line of sight as it passes through the disk, and, 
consequently, the observed rotation measure would be
greatly reduced.   
Future VLBI polarization observations may be able to
directly measure the Faraday rotation and resulting
depolarization in front of the radio counterjet.

\section{Conclusions} 

Our new observations at 1.6 and 4.9 GHz confirm the free-free
absorption explanation for the sub-parsec radio morphology in 
NGC 4261, specifically the fact that the gap in emission at
the base of the counterjet is caused by shadowing of the
counterjet by the accretion disk.    
Measurements of the angular width and optical depth 
in the absorbed region, 
and the distance between the absorption and the core, as a function
of frequency show that the inner pc of the accretion disk is
geometrically thin and that the  
average electron density is a few times $10^{4} \ {\rm cm}^{-3}$
at a deprojected radius of 0.2 pc.  The corresponding equipartition
magnetic field strength is about $10^{-4}$ G.  
Similar observations of other galaxies covering a wide range
of radio luminosities will allow correlations between the 
central engine fueling process (accretion disk geometry, density,  
mass, and magnetic field) and AGN activity level (nonthermal luminosity)
to be studied.  Such studies should improve our understanding
of the physical processes involved in accretion-powered central
engines in galactic nuclei.  

\acknowledgements
We gratefully acknowledge the VSOP Project, which is led
by the Japanese Institute of Space and Astronautical
Science in cooperation with many organizations and radio
telescopes around the world.  
We also thank the anonymous referee for helpful comments. 
AEW~acknowledges support from the NASA Long Term 
Space Astrophysics Program.
This research was carried out at the Jet 
Propulsion Laboratory, California Institute of Technology, under 
contract with the National Aeronautics and Space Administration.

\end{document}